\documentclass[12pt,fleqn]{article}
\usepackage{espcrc1}

\usepackage{graphicx}
\usepackage[figuresright]{rotating}

\newcommand{\AmS}{{\protect\the\textfont2
  A\kern-.1667em\lower.5ex\hbox{M}\kern-.125emS}}

\title{Evidence for a long range structure in the pion emission source in Au+Au collisions at RHIC}

\author{P. Chung\address[sunysb]{Dept. of Chemistry, 
        SUNY Stony Brook, Stony Brook, NY 11794, USA}, %
        for the PHENIX Collaboration
       }
       
\begin{document}

\maketitle

\begin{abstract}
The PHENIX experiment has recently acquired $\sim$ 1 billion minimum bias Au+Au events 
at $\sqrt s = 200$AGeV during the year-2004 run. This high statistics data set, 
coupled with a state-of-the-art analysis technique, allows for the extraction of 3D emission 
sources for various particle types. These 3D sources lend fresh insight into 
the nature of a long-range source previously reported by PHENIX. The new results indicate  
an anisotropic pion emission source in the pair center of mass system (PCMS) having an extended space-time extent 
oriented in the outward direction. The two-proton emission source from the same data set 
is essentially isotropic in the PCMS. These results provide 
a ``window'' for viewing the evolution dynamics of the high energy density nuclear matter 
created at RHIC.
\end{abstract}

\section{Introduction}

 A de-confined phase of nuclear matter is expected to be formed at the high energy densities 
 created in Au+Au collisions at RHIC\cite{qgp03}. 
An observation of the presence (or absence) of an emission source of large space-time extent 
for particle emission can provide important constraints for understanding the nature of 
this phase transition.

In recent measurements, the PHENIX Collaboration has observed a long range structure in the
1D two-pion emission source function for Au+Au collisions at $\sqrt s=200$ AGeV 
\cite{chung05}. This long range structure was resolved using the 1D Source Imaging technique 
of Brown and Danielewicz\cite{brown97,brown98}. The 1D technique does not give crucial 
directional information so a detailed analysis of the dynamical origin of this structure was not 
possible. To address this shortcoming, Danielewicz and Pratt introduced the more powerful 
technique of decomposing correlation functions into correlation moments using a 
cartesian surface-spherical harmonics basis \cite{daniel05}. In this representation, 
each moment (of a particular order) corresponds to a specific deformation 
( i.e dipole, quadrupole etc ) of the 3D source function, and their extraction provide 
detailed 3D information about the emission source. 

In this paper, we present the first application of the moment decomposition
technique for the analysis of $\pi^+\pi^+$ and pp pairs produced in Au+Au collisions at
$\sqrt s=200$ AGeV.

\section{Experimental Setup and Data Analysis}

The data presented here were taken by the PHENIX Collaboration during the year-2004 run. The 
colliding beams ($\sqrt s = 200$ AGeV) were provided by the RHIC accelerator.
Charged tracks were detected in the two central arms of PHENIX 
\cite{phenix03}, each of which subtends 90 degrees in azimuth and $\pm$0.35 
units of pseudo-rapidity. Tracking information was provided by a drift 
chamber followed by two layers of pad chambers. Particle identification was 
performed by an electromagnetic calorimeter and a time-of-flight wall.


3D correlation functions , C($\mathbf{q}$), were obtained as the ratio of foreground
to background distributions in relative momentum $\mathbf{q}$ for $\pi^+\pi^+$ and pp pairs. Here, $\mathbf{q}=\frac{(\mathbf{p_1}-\mathbf{p_2})}{2}$ is half of the relative
momentum between the two particles in the PCMS frame. The foreground
distribution was obtained using pairs of particles from the same event and the background 
was obtained by pairing particles from different events. The events used have a z-vertex position within $\pm30$cm from the center of the PHENIX spectrometer. Track 
merging and splitting effects were removed by appropriate cuts in the
relevant coordinate space on both the foreground and background distributions. 
There was no significant effect on the correlation functions due to the
momentum resolution of 0.7$\%$. 

In the cartesian harmonic decomposition, the 3D correlation function is expressed as
\begin{equation}
 C(\mathbf{q}) = \sum_l \sum_{\alpha_1 \ldots \alpha_l} 
   C^l_{\alpha_1 \ldots \alpha_l}(q) \,A^l_{\alpha_1 \ldots \alpha_l} (\Omega_\mathbf{q})
\end{equation}
where $l=0,1,2,\ldots$, $\alpha_i=x, y \mbox{ or } z$, $A^l_{\alpha_1 \ldots \alpha_l}(\Omega_\mathbf{q})$ 
are cartesian harmonic basis elements ($\Omega_\mathbf{q}$ is solid angle in $\mathbf{q}$ space) and $C^l_{\alpha_1 \ldots \alpha_l}(q)$ are cartesian correlation moments given by
\begin{equation}
 C^l_{\alpha_1 \ldots \alpha_l}(q) = \frac{(2l+1)!!}{l!} 
 \int \frac{d \Omega_\mathbf{q}}{4\pi} A^l_{\alpha_1 \ldots \alpha_l} (\Omega_\mathbf{q}) \, C(\mathbf{q})
\end{equation}
The cartesian coordinate system is oriented such that z is parallel to the beam (longitudinal), x points 
in the direction of the total momentum of the pair in the PCMS frame (outward) and y is perpendicular 
to the other two axes (sidewards).

\section{Results}

\begin{figure}

\begin{minipage}[t]{0.5\linewidth}
\includegraphics[width=1.\linewidth]{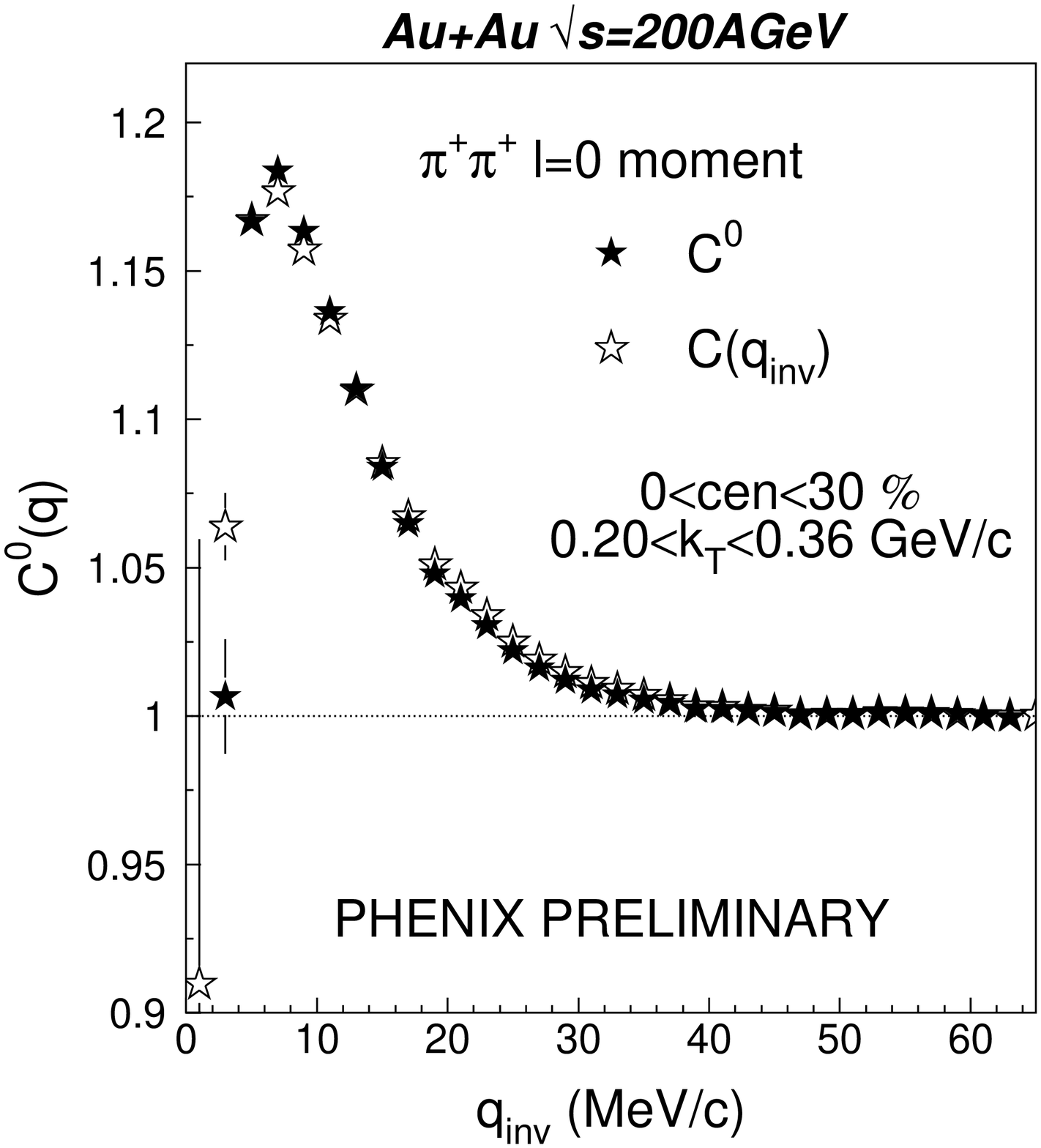}
\vskip -1.2cm
\caption{\small{ $\pi^+\pi^+$ l=0 moment $C^0$ and 1D correlation C($q_{inv}$) for centrality 0-30$\%$ and $0.20<k_T<0.36$ GeV/c in Au+Au collisions.}}
\label{pipi_c0}
\end{minipage}
\hskip 0.2cm
\begin{minipage}[t]{0.5\linewidth}
\includegraphics[width=1\linewidth]{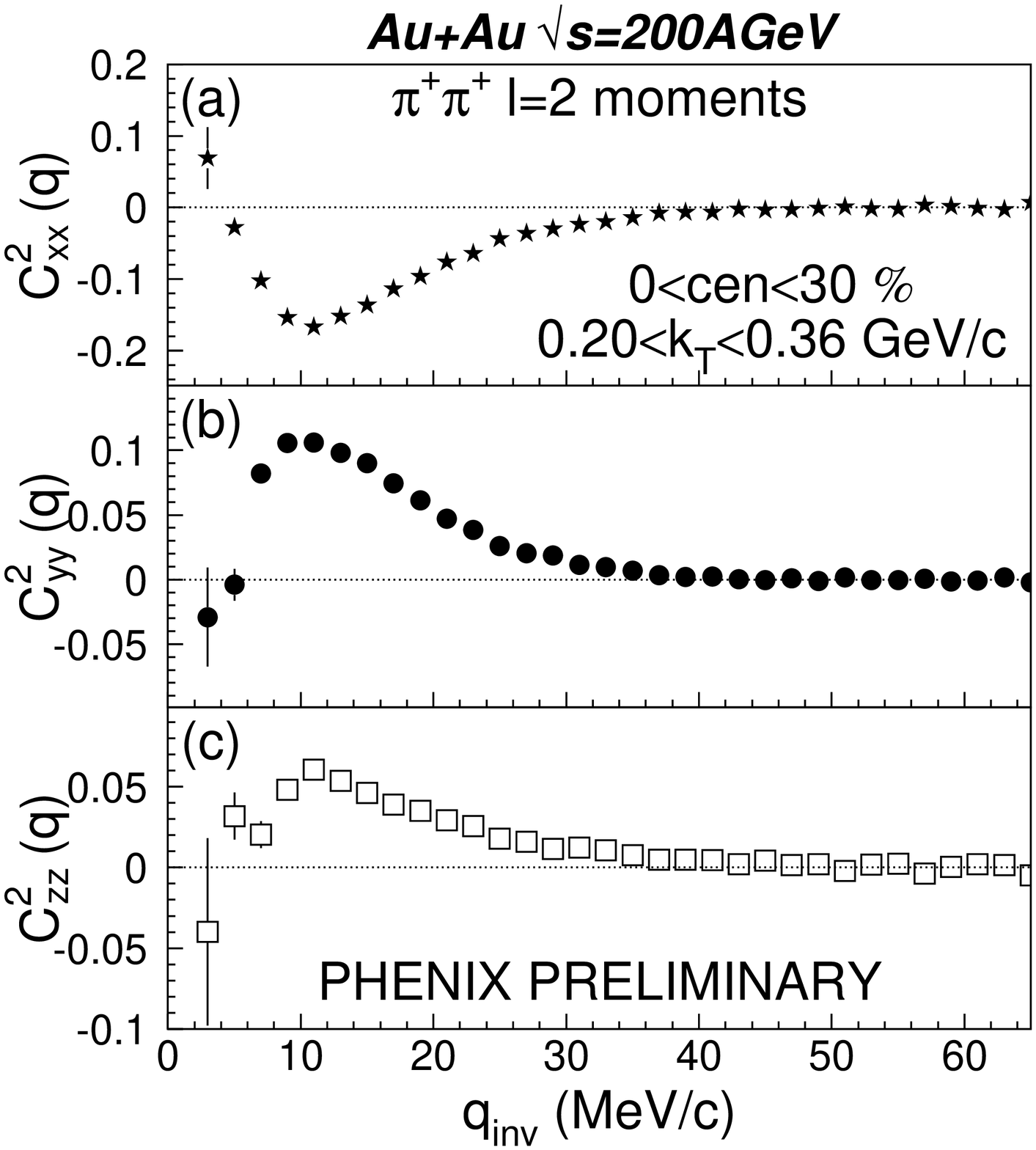}
\vskip -1.2cm
\caption{\small{$\pi^+\pi^+$ l=2 moments (a) $C^2_{xx}$ (b) $C^2_{yy}$ and (c) $C^2_{zz}$ for centrality 0-30$\%$ and $0.20<k_T<0.36$ GeV/c in Au+Au collisions.}}
\label{pipi_c2}
\end{minipage}

\end{figure}

Fig.~\ref{pipi_c0} shows the l=0 moment $C^0$ (solid stars) and 1D correlation function C(q) (open stars) 
for $\pi^+\pi^+$ pairs with $0.20<k_T<0.36$ GeV/c, from Au+Au collisions in the centrality range 0-30$\%$ 
of total cross section. Here, $k_T = \frac{ (\mathbf{p_{1T}} + \mathbf{p_{2T}}) }{2}$ is the mean tranverse momentum of the two particles in the particle pair.
The l=0 moment is essentially identical to the 1D correlation function as 
expected. We conclude that fig.~\ref{pipi_c0} serves as a good consistency check of the moment 
calculation procedure.

Figs.~\ref{pipi_c2}(a),(b) and (c) show the l=2 moments $C^2_{xx}$, $C^2_{yy}$ and $C^2_{zz}$ respectively 
for $\pi^+\pi^+$ pairs with $0.20<k_T<0.36$ GeV/c, from Au+Au collisions in the centrality range 0-30$\%$. 
The non-zero values for the l=2 moments represent source anisotropies of a quadrupole nature and point to 
specific deformations in the direction specified by the moment. More specifically, the fact that 
$C^2_{xx}$ is negative for all $q_{inv}$ values indicates a negative contribution to the correlation 
function in the x (outward) direction. This is to be compared to the results for $C^2_{yy}$ and $C^2_{zz}$ 
which are both positive and hence represent positive contributions to the correlation function 
in the y (sideward) and z (longitudinal) directions. Thus, the overall correlation function, 
obtained by adding the l=0 and l=2 moments, is narrower in x and broader in the y and z directions, indicating the 3D
source size is larger in x and smaller in y and z directions, as
compared to the angle-averaged source size.
 
\begin{figure}

\begin{minipage}[t]{0.46\linewidth}
\includegraphics[width=1.\linewidth]{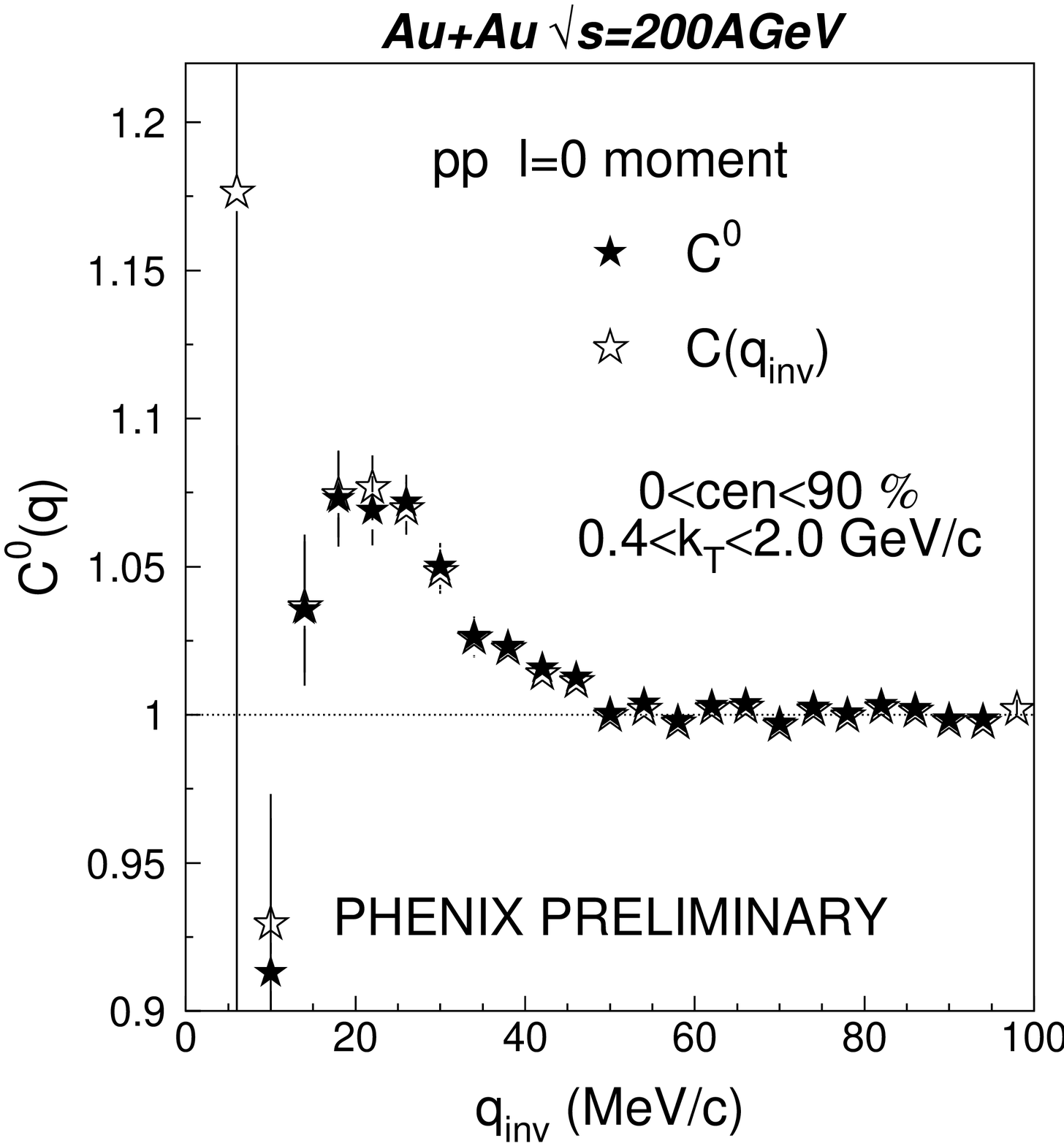}
\vskip -1.2cm
\caption{\small{ pp l=0 moment $C^0$ and 1D correlation C($q_{inv}$) for centrality 0-90$\%$ and $0.4<k_T<2.0$ GeV/c in Au+Au collisions.}}
\label{pp_c0}
\end{minipage}
\hskip 0.2cm
\begin{minipage}[t]{0.46\linewidth}
\includegraphics[width=1.\linewidth]{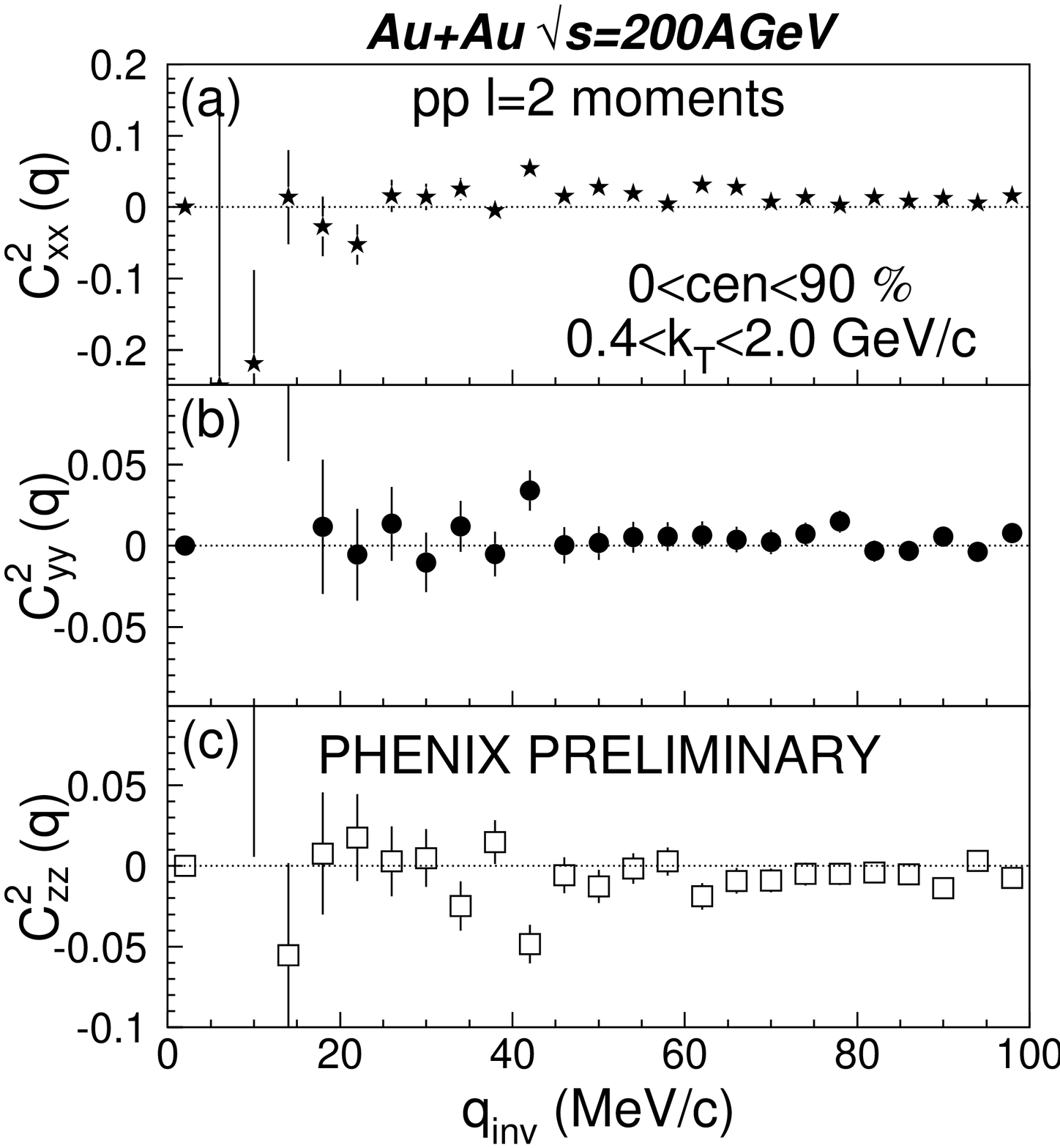}
\vskip -1.2cm
\caption{\small{pp l=2 moments (a) $C^2_{xx}$ (b) $C^2_{yy}$ and (c) $C^2_{zz}$ for centrality 0-90$\%$ and $0.4<k_T<2.0$ GeV/c in Au+Au collisions.}}
\label{pp_c2}

\end{minipage}

\end{figure}

Fig.~\ref{pp_c0} shows the l=0 moment $C^0$ (solid stars) and 1D correlation function C(q) (open stars) 
for pp pairs for $0.4<k_T<2.0$ GeV/c, from Au+Au collisions in the centrality range 0-90$\%$. Again, the very
good agreement between the l=0 moment and the 1D correlation function attests to the reliability of the 
moment calculation procedure.

Figs.~\ref{pp_c2}(a),(b) and (c) show the l=2 moments $C^2_{xx}$, $C^2_{yy}$ and $C^2_{zz}$ 
respectively for pp pairs, with $0.4<k_T<2.0$ GeV/c, from Au+Au collisions in the centrality 
range 0-90$\%$. In contrast to the $\pi^+\pi^+$ moments, the l=2 moments for the pp correlation function are
all consistent with 0 within statistical fluctuations. Hence, the overall pp
correlation function is the same in all 3 directions, namely the angle-averaged
l=0 moment $C^0$. 

\section{Discussion}

Analysis of the $\pi^+\pi^+$ 1D correlation function indicates a long range structure in the angle-averaged 
pion source function \cite{chung05}. The cartesian moments, shown in Fig.~\ref{pipi_c2}, indicate that the 
correlation function is narrower in the outward direction compared to the sideward and longitudinal 
directions. The consequence of this is a source function which is more elongated in the outward direction in the PCMS frame.

This elongation must necessarily be due to prolonged emissions in the PCMS. A source isotropic in its rest 
frame, but breaking up instantaneously in the PCMS frame, would have been shortened in the outward 
direction by Lorentz factor $\gamma_{boost}$. On the other hand, a source breaking up instantaneously in 
its own rest frame would lead to a source elongated in the PCMS by the factor $\gamma_{boost}$ in the 
outward direction. Here, non-simultaneity of emission and movement of the source would lead to 
overcompensation of the Lorentz contraction. The most extreme Lorentz elongation factor would be obtained under the 
assumption of a source freezing out instantaneously in the locally co-moving frame. However, such a  
source would be unlikely. Instead, a more plausible source might be one with a radial velocity 
along the pair momentum, that freezes out over a finite time in its frame. 
Further conclusions await the results of detailed model studies with global and/or dynamic 
parametrization of the emission.

 In contrast to the pions, the cartesian moments for pp pairs, shown in
Figs.~\ref{pp_c0} and \ref{pp_c2}, indicate that the proton source function is 
isotropic in the PCMS. This suggests a proton emission source having a velocity close 
to that for the pair velocity.
 
 The current results constitute an important model constraint because a basic requirement will be 
to explain the pion and proton emission differences and the associated correlation anisotropies
simultaneously.

\section{Acknowledgements}
The authors are grateful to Drs. P. Danielewicz and S. Pratt for fruitful discussions.

\end{document}